\documentclass[12pt]{article}

\usepackage{pstricks}
\usepackage{graphicx}
\usepackage{graphics}
\usepackage{amssymb}
\usepackage{epsfig,caption,graphicx,eepic,epic}
\usepackage{pstricks}
\usepackage{amsmath}
\usepackage{multido}

\begin{document}
\title{Propagation of local excitations through strongly correlated quantum chains \\ } 

\author{Jean Richert$^{a}$
\footnote{E-mail address: richert@fresnel.u-strasbg.fr}\\
and\\
Tarek Khalil$^{b,c}$
\footnote{E-mail address: tarek.khalil@liu.edu.lb}\\
$^{a}$ Institut de Physique, Universit\'e de Strasbourg,\\
3, rue de l'Universit\'e, 67084 Strasbourg Cedex,\\ 
France\\
$^{b}$ Department of Physics, School of Arts and Sciences,\\
Lebanese International University, Beirut, Lebanon\\
$^{c}$ Department of Physics, Faculty of Sciences(V),\\
Lebanese University, Nabatieh, Lebanon} 

\date{\today}
\maketitle 

\begin{abstract}
The propagation of an external transverse magnetic signal acting locally on a 1$d$ chain of spins 
generates a disturbance which runs through the system. This quantum effect can be interpreted as 
a classical traveling wave which contains a superposition of a large set of frequencies depending on the 
size of the chain. Its local amplitude fixes the size of the z-component of the spins at any location 
in the chain. The average and maximum value of the group velocity are determined and compared with the 
transmission velocity fixed by the Lieb-Robinson upper bound inequality.
\end{abstract} 

\maketitle


\section {Introduction}

Modern communication needs the development of technologies which allow fast, faithful and robust 
information manipulation by means of adapted transfer devices. These needs generate intense research 
in different fields of condensed matter physics in order to develop efficient material tools on 
which information can be safely propagated. 

Indeed, the last decade saw the development of new several techniques and methods in the field of 
quantum information technology such as ultra-fast and flexible optical manipulations of spins in 
magnetically ordered materials~\cite{ban,len,rub}, transfer of quantum states and qubits along linear 
systems and different techniques~\cite{alb,bos,chr1,chr2,heu,bur1,wan,shi,bur2,pem,cap}, superconducting 
devices~\cite{rom,lya}, transfer of entangled states~\cite{ver,sub,ple1,ple2,mar}, investigations 
concerning echoes and decay times~\cite{gou}, transition times~\cite{mur,jon}, propagation 
velocities in solid state systems~\cite{lr1,lr2,lr3,lr4,ham,isa,sch,pou}. The subject raised also more 
fundamental points concerning the properties related to the non-equilibrium behaviour of such systems
~\cite{igl,osb,hap}. 

The present work uses a simple model, a finite $XY$ chain submitted to even boundary conditions in its ground 
state. It works here as the material support for the propagation of external perturbations along the ring chain. 
The perturbations are chosen to be local excitations induced on a single spin over a short interval of time 
and the propagation of the signal in time is followed over the whole chain. The interest here lies essentially on 
the estimate of the propagation velocity of a signal through the chain. The perturbation behaves like 
a traveling wave characterized by a set of frequencies which depend on the size and eigenenergies of a
strongly correlated system. This quantity can be obtained in an analytic form and group velocities are defined. 
Their order of magnitude are obtained by means of numerical applications and compared with the upper velocity 
obtained in the application of the Lieb-Robinson (LR) approach to the transmission of information in discrete 
quantum systems.

 Section $2$ introduces the model and the time evolution of the z-component of local spins along the chain.
In section $3$ different group velocities are defined and the expression of the LR bound velocity is recalled.
Numerical applications are presented in section $4$ and the results concerning the propagation are discussed.
Section $5$ is concerned with some considerations concerning the control of the local spin amplitudes during the 
propagation process and the implementation of sequences of local excitations in the system. Conclusions are drawn 
in section $6$.

\section {The physical system}

\subsection {The model}
 
The Hamiltonian of the finite $1d$ quantum chain with $N$ sites and asymmetry parameter $\gamma$ reads
 
\begin{eqnarray}
H_{0} = J/2 (1+\gamma)\sum_{(i)} \sigma^{x}_{i}\sigma^{x}_{i+1} + J/2 (1-\gamma)\sum_{(i)} \sigma^{y}_{i}\sigma^{y}_{i+1}
\notag \\ 
- h_{0}\sum_{(i)}\sigma^{z}_{i}
\label{eq1}   
\end{eqnarray} 
In the sequel even periodic boundary conditions are introduced. The interaction strength is chosen 
to be $J=1$ and $\hbar=1$. In Fourier space each site $[k=1,...,N/2]$ contains four single spin states. The energy 
spectrum can be determined analytically in terms of single particle energies~\cite{bar,kat}. Diagonalization of 
the single particle ($4*4$) Hamiltonian leads to the stationary orthonormalized eigenstates 
\begin{eqnarray}
|\psi^{1}_{k}\rangle &=& \alpha^{k}_{1}|0\rangle_{k} + \alpha^{k}_{2} a^{+}_{k}a^{+} _{-k}|0\rangle_{k}
\notag \\
|\psi^{2}_{k}\rangle &=& \beta^{k}_{1}|0\rangle_{k} + \beta^{k}_{2} a^{+}_{k}a^{+}_{-k}|0\rangle_{k}
\notag \\
|\psi^{3}_{k}\rangle &=&a^{+}_{k}|0\rangle_{k}
\notag \\
|\psi^{4}_{k}\rangle &=& a^{+}_{-k}|0\rangle_{k}
\label{eq2}   
\end{eqnarray} 
where $|0\rangle_{k}$ is the spin vacuum, $a^{+}_{k}$ creates a spin at site $k$, $\alpha^{k}$ and 
$\beta^{k}$ are complex amplitudes which are shown in Appendix A.

The corresponding eigenvalues read

\begin{eqnarray}
\epsilon^{(1)}_{k} &=& \cos\phi_{k}-[(\cos\phi_{k}- h_{0})^{2} + \gamma^{2} \sin^{2}\phi_{k}]^{1/2} 
\notag \\
\epsilon^{(2)}_{k} &=& \cos\phi_{k}+[(\cos\phi_{k}- h_{0})^{2} + \gamma^{2} \sin^{2}\phi_{k}]^{1/2} 
\notag \\
\epsilon^{(3)}_{k} &=& \cos\phi_{k}=\epsilon^{(4)}_{k} 
\label{eq3}
\end{eqnarray} 
where $\phi_{k} = 2 \pi k/N$.

The $Nth$ spin of the system experiences an external magnetic field $h_{1}$ which induces a perturbation 
\begin{eqnarray}
H^{(N)}_{1}(t)= h_{1}\exp(-t/\tau_{H}) S^{z}_{N}
\label{eq4}   
\end{eqnarray} 
with $S^{z}_{N}=\sigma^{z}_{N}/2$. The external field  acts on the $z$ component of the spin at site $N$ 
over a finite time interval fixed by $\tau_{H}$ and starting at $t=0$. The generated signal is transmitted 
to the whole chain of connected interacting spins. The localized action of an external magnetic field 
may be an experimental challenge. There exists however experimental work which shows that this might be 
possible in the not too far future~\cite{luk,ram,ron}.

\subsection {Time evolution of the individual spins in the chain}

The $z$ component of each spin site $n$ evolves in time as  $S^{z}_{n}(t)$
\begin{eqnarray}
S^{z}_{n}(t)=\exp(i \int_{0}^{t}H(t')dt')S^{z}_{n}(0)\exp(-i \int_{0}^{t}H(t')dt')
\label{eq5}   
\end{eqnarray}
where $H(t)=H_{0}+H^{(N)}_{1}(t)$.

The physical situation considered here corresponds to an excitation induced by $H^{(N)}_{1}(t)$ which relaxes 
over a time $\tau_{H}$. This time is chosen as small compared to the time unit. In this regime a perturbative treatment 
of the expectation value of $S^{z}_{n}(t)$ makes sense. The perturbation is essentially governed by the parameter 
$\tau_{H}$ and $h_{1}\tau_{H}$ which can be small in practical applications if $\tau_{H}$ is a very short 
relaxation time even for sizable magnetic fields $h_{1}$.

Using a perturbative treatment up to second order in the perturbation development provides an analytic expression 
of the time evolution of the spins along the chain. The ground state $|\Psi_{0}\rangle$ of the system is the product 
of the single particle states. A second order perturbation expansion gives the expectation value $\langle S^{z}_{n}(t)\rangle$ 
of the z-component of the spin with respect to the ground state delivers three contributions

\begin{eqnarray}
\langle S^{z}_{n}(t)\rangle=\langle S^{z}_{n}\rangle^{(0)}+\langle S^{z}_{n}(t)\rangle^{(1)} +\langle S^{z}_{n}(t)\rangle^{(2)}
\label{eq6}   
\end{eqnarray}
 
where 

\begin{eqnarray}
\langle S^{z}_{n}\rangle^{(0)}=2/N \sum_{k=1}^{N/2} |\alpha^{k}_{2}|^{2}-1/2
\label{eq7}   
\end{eqnarray}

\begin{eqnarray}
\langle S^{z}_{n}(t)\rangle^{(1)}=1/2\sum_{k,l=1}^{N/2}v_{2}^{(k,l)}[A_{kl}[\sin(\psi_{n;kl}^{+})
+\sin(\psi_{n;kl}^{-})]
\notag \\ 
+ B_{kl}[\cos(\psi_{n;kl}^{+})+\cos(\psi_{n;kl}^{-})]]+C(t,\tau_{H})
\label{eq8}   
\end{eqnarray}
where $v_{2}^{(k,l)}$ are transition matrix elements between different single spin states ($k,l$), $A_{kl}$ and $B_{kl}$
amplitudes, $C(t,\tau_{H})$ a time-dependent and state independent contribution which relaxes to a constant over a 
time interval $t \ge \tau_{H}$. The phases read
\begin{eqnarray}
\psi_{n;kl}^{+}= \omega_{(kl)}t+n K(k,l)
\notag \\
\psi_{n;kl}^{-}= \omega_{(kl)}t-n K(k,l)
\label{eq9}   
\end{eqnarray}          
with $\omega_{(kl)}=\epsilon^{(3)}_{k}+\epsilon^{(4)}_{l}-\epsilon^{(1)}_{k}-\epsilon^{(1)}_{l}$
and $K(k,l)=(\phi_{k}-\phi_{l})$. The expressions of the matrix elements $v_{2}^{(k,l)}$
and the amplitudes $A_{kl}$ and $B_{kl}$ are shown in Appendix B.

The expression of the second order contribution $\langle S^{z}_{n}(t)\rangle^{(2)}$ is given by a large number of terms. 
Present applications show that their contributions are sizably smaller than those of the first order when $t\gg\tau_{H}$. 
In Appendix C we estimate their order of magnitude. 

\section {Group velocities}

The time-dependent contribution of $\langle S^{z}_{n}(t)\rangle^{(1)}$ given by equation (8) can be interpreted
as a sum of classical traveling waves characterized by frequencies $\omega_{(kl)}$ and phases $(\phi_{k}-\phi_{l})$. 
Considering the analytic expressions of the $\omega_{(kl)}$ and the energies ${\epsilon^{(i)}_{k}}$ given above one sees 
that these quantities are incommensurate. Hence the oscillations will not be periodic in time. The group velocity for a 
fixed wave number $K(k,l)$ characterizes the propagation of the waves. This quantity is defined as 
\begin{eqnarray}
v_{gr}=d\omega(K)/dK
\label{eq10}
\end{eqnarray}

For a fixed value of $K(k,l)$ more than one value of $\omega(k,l)$ is possible since $K(k,l)$ depends on the difference $k-l$. 
In the sequel we consider $\omega^{max}(K)$ which correponds to maximum values of $\omega$s for a given set of couples $(k,l)$ 
corresponding to a fixed $K$ and the averages $<\omega>$ over all possible couples $(k,l)$ generating the same $K$. 

The group velocity is a measure of the speed at which the external perturbation propagates through the medium, here in the 
case of a spin chain. By means of Eq.(10) it is possible to define and obtain an order of magnitude of this quantity and an 
upper value of the propagation of a signal in this type of quantum devices. Explicit calculations are presented and discussed below. 

On the other hand the Lieb-Robinson procedure~\cite{lr1} allows the determination of an upper limit of the velocity $v$ with
which a signal is transmitted in a discrete system from an initial point at a fixed initial time to a final point at a fixed later time
~\cite{lr2,lr3,lr4}. This velocity  depends on metric factors which define distances between sites of the system and the 
norm $\|H\|$ of the Hamiltonian matrix. The limit velocity is obtained by means of a variational procedure depending on a real positive 
parameter $a$. Working out the expression and determining the upper bound by varying $a$ leads to the expression
\begin{eqnarray}
v_{LR}=eN\|H\|/2
\end{eqnarray}
where $e$ is the exponential constant, $N$ the number of sites of the system. Some details of the derivation are given in 
Appendix D. In section $4.3$ we work out $v_{LR} $ and compare it with $v_{gr}$.

\section {Explicit calculations}
 
\subsection{Preliminaries}

The time evolution of a fixed spin and the behaviour of the spins in the chain are worked out for a spin chain 
of finite length. The physical conditions under which the perturbative treatment discussed above is valid are
realized in the present applications. The zeroth order contribution to the z-component of the spin 
$\langle S^{z}_{n}\rangle^{(0)}$ is time-independent and the second order term $\langle S^{z}_{n}\rangle^{(2)}$
is smaller than $\langle S^{z}_{n}\rangle^{(1)}$ by two orders of magnitude. Due to the symmetries induced by the 
boundary conditions imposed on the chain the oscillation frequencies $\omega(k,l)$ are symmetric with respect 
to $k$ and $l$ belonging to the intervals $[1,N/2],[-1,-N/2]$. Hence the calculations can be restricted to the 
interval $[1,N/2]$. In the sequel we consider a finite chain of length $N=100$ and a set of parameters 
$(h_{0}=0.5,h_{1}=1.0, \gamma=0.5$ which correspond to a system which lies above the critical point 
$h_{0}=h_{c}=-0.5$ after the action of the short excitation corresponding to $\tau_{H}=10^{-4}$.


\begin{figure}
\epsfig{file=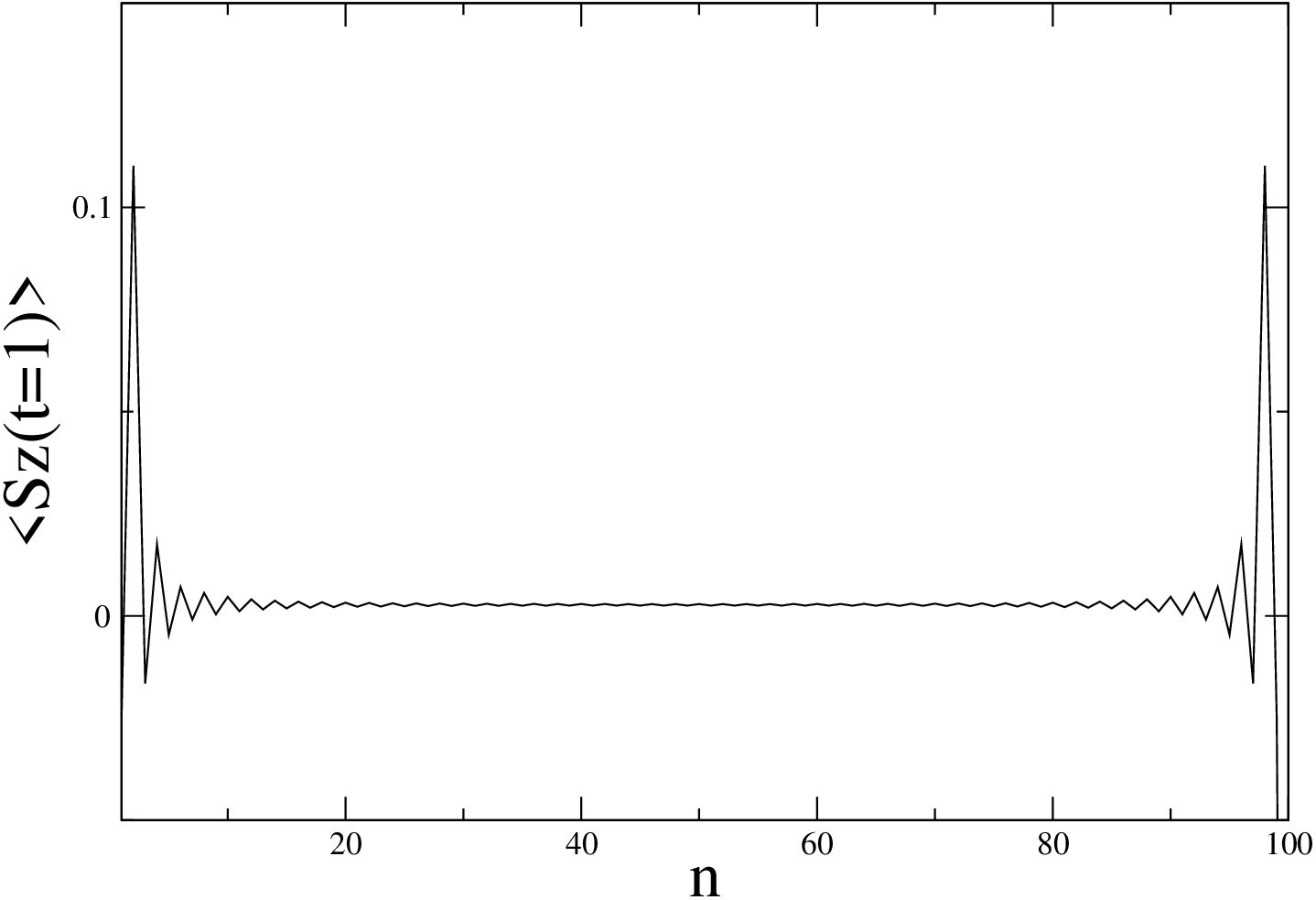,scale=0.30}
\caption{z-components of the spins $n$ for t=1.}
\label{fig1}
\end{figure}

\subsection{Spin oscillations along the chain}

 The time evolution of 
\begin{eqnarray}
\langle S^{z}_{n}(t)\rangle^{(12)}=\langle S^{z}_{n}(t)\rangle^{(1)}+\langle S^{z}_{n}(t)\rangle^{(2)}
\label{eq19}   
\end{eqnarray} 
for fixed time and a set of sites $n=1$ to $50$ is shown in Figs. $1$ and $2$.

For short times the essential part of the excitation resides in the sites with low $n$ which are close to $N=100$ due to the
periodic boundary conditions. For large times the longitudinal part of all the spins is affected and shows  
a somewhat irregular and non repetitive pattern due to the incommensurability of the values of $\omega$ and $K(k,l)$ discussed 
above. The calculations have been repeated for an excitation $h_{1}>0$ above $h_{0}=h_{c}=-1/2$ where the system experiences 
a phase transition as well as an excitation which leads to a crossing of the critical point when $t$ goes to infinity. In both 
cases no specific behaviour of the spin component is observed. Individual spins which are affected by local excitations are qualitatively insensitive to the global effects induced phase transitions which affect the whole system.

\begin{figure}
\includegraphics[scale=0.30]{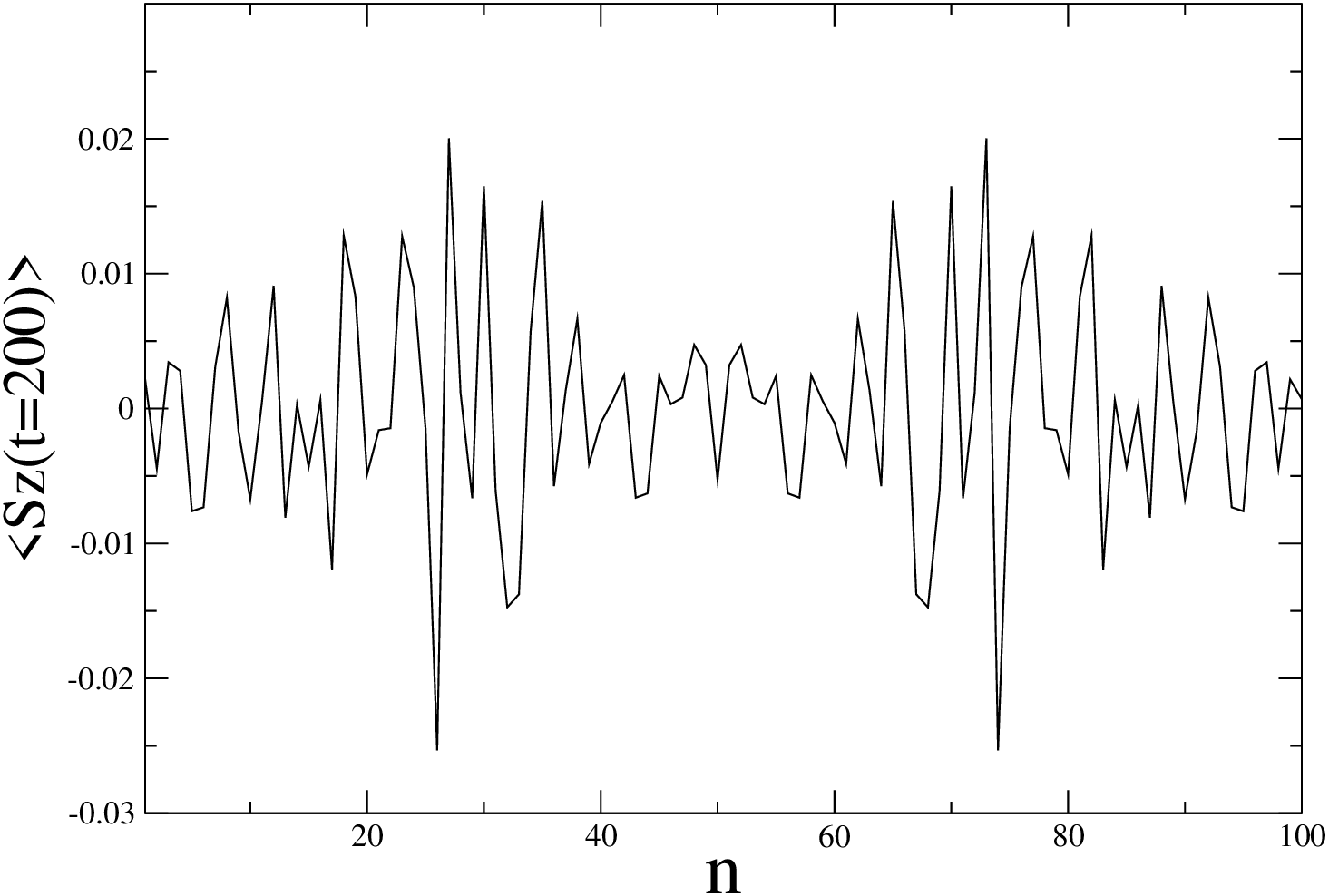}
\caption{z-components of the spins $n$ for t=200.}
\label{fig2}
\end{figure}

Figs. $3$ and $4$ show the behaviour of $\langle S^{z}_{n=1}(t)\rangle^{(12)}$ and $\langle S^{z}_{n=40}(t)\rangle^{(12)}$
over an interval of time $t=[1,200]$. It is seen that the evolution of the spins in time corresponds globally to a wave
which spreads over all spins and decreases necessarily in amplitude since the total magnetization of the system is constant 
and the perturbation is small. 

\subsection{Group velocities}
 
Figs. $(5)$ and $(6)$ show the values of the maximal and average group velocities $v^{max}_{gr}(K)$ and $< v_{gr}(K)>$ for fixed $K=[1,N/2-1]$, $K(k,l)=|k-l|$. The parameters of the model are those given in section $4.1$. One observes sizable variations of these quantities over the range of $K$ values which shows a large dispersive behaviour of the system due to the fact that the frequencies show a strongly non-linear dependence on $K(k,l)$. This quantity  is is itself a mixture of many single particle energies. The absolute value of the maxima are 1.5 for $v^{max}_{gr}(K)$ and 0.55 for $< v_{gr}(K)>$ which correspond to a maximum transit time between two sites of 0.66 time units for $v^{max}_{gr}(K)$ and 1.8 for $< v_{gr}(K)>$. These quantities are independent of the amplitude of $h_{1}$, the velocities depend on the characteristics of the system only.
 
\vskip 0.3cm

Using the expression of the Lieb-Robinson velocity for a signal which evolves over the {\it whole} system one finds $v_{LR}=1.02*10^4$ ($N=100$ and $\|H\|=75$) in the same physical units  as those used for $v_{gr}$. It appears that there is a large quantitative difference between the two velocities. One can make the following comments. 

The group velocity $v_{gr}$ generated by an external disturbance is a quantity which is local in $K$ space and depends on single particle energy differences, see section 2.2. On the other hand $v_{LR}$ is the maximum speed with which a signal emitted at some time at some place is transmitted through the system and observed at some other place at a later time. It depends explicitly on the extensive properties of the system, the maximum of its total energy ($\|H\|$) and its size ($N$). The transmission between two neighbouring sites ($N=2$) one would lead to $v_{LR}=4.07$.

The expression of $v_{LR}$ depends on the geometric stucture of the system only through its total energy ($\|H\|$) which although different should be of the same order of magnitude for open and closed systems. This effect is certainly weaker than the difference between $v_{gr}$ in a closed and an open system due to the wave nature of the propagation.

\section {Information transmission along the chain}

\subsection {Spin amplitude transportation}

Consider the transmission of the amplitude of the z-component $\langle S^{z}_{N}(t)\rangle^{(1)}$ located at 
site $N$ at time $t$ to some other site in the chain. This may be realized in the following way.

At a given initial time $t$ and fixed $h_{1}$ and $\tau_{H}$ one applies an excitation at site $N$ which generates a 
perturbation $\langle S^{z}_{N}(t)\rangle^{(1)}$. At time $t + \Delta t$, $\langle S^{z}_{N-1}(t+\Delta t,h_{1})\rangle^{(1)}$ 
has a well defined value. If one applies a further excitation $h'_{1}$ at site $N-1$ and time $t + \Delta t$ such that  

\begin{eqnarray}
\langle S^{z}_{N}(t)\rangle^{(1)}=\langle S^{z}_{N-1}(t+\Delta t,h_{1})\rangle^{(1)} + 
\notag \\\
\langle S^{z}_{N-1}(t+\Delta t,h'_{1})\rangle^{(1)}
\label{eq13}
\end{eqnarray}
the amplitude of the spin component at site $N-1$ at time $t+\Delta t$ will be the same as the amplitude at site $N$
at time $t$. The field $h'_{1}$ needed is easily fixed since any z-component of a spin
$\langle S^{z(1)}_{n}(t+\Delta t,h_{1})\rangle$ depends linearly on the field $h_{1}$ under the physical conditions 
described above. The operation can be iterated over a finite interval of sites of a given length. 


\begin{figure}
\includegraphics[scale=0.30]{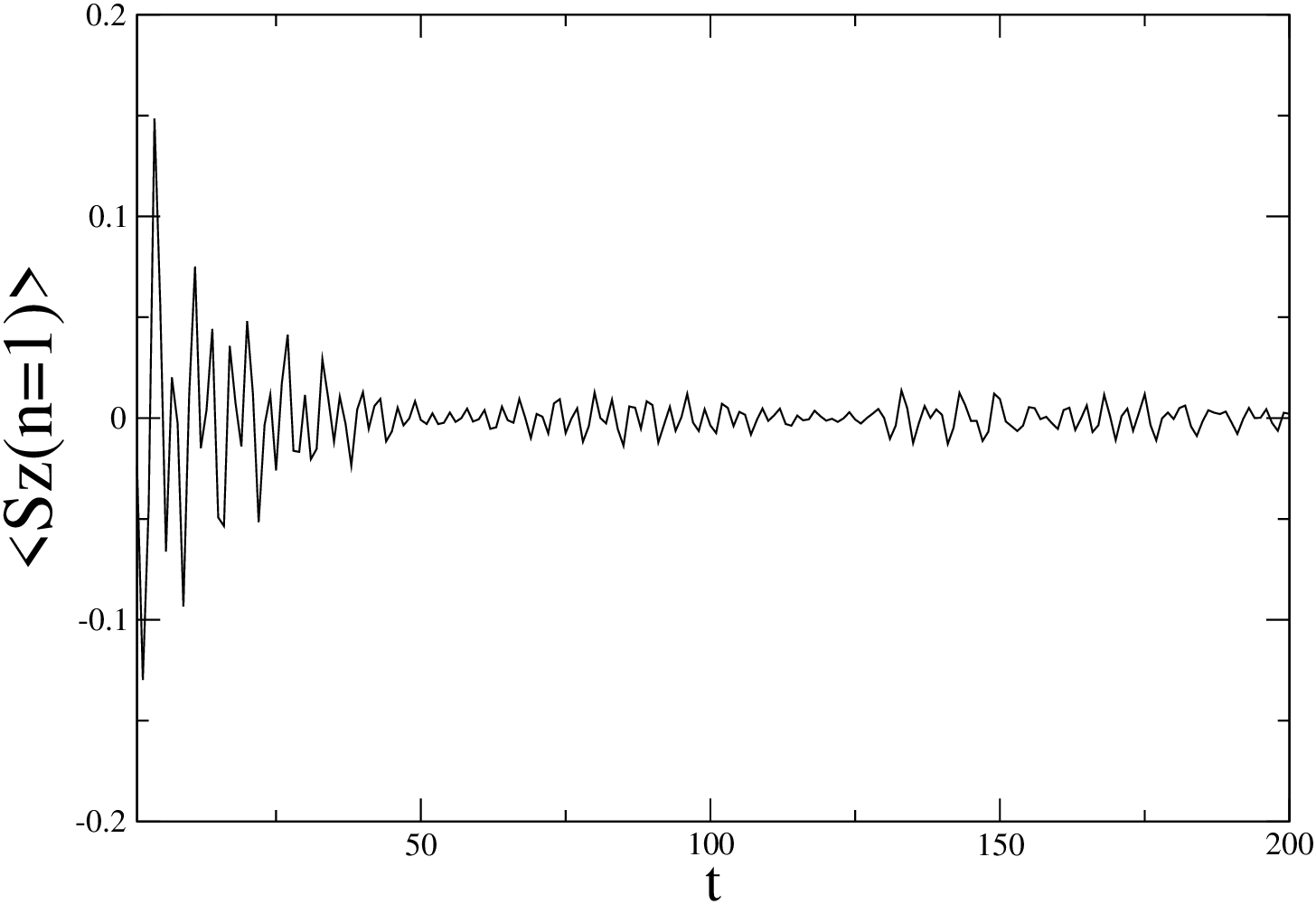}
\caption{Time dependence of the z-component $\langle S^{z}_{n}(t)\rangle^{(12)}$ of the spin at site n=1.}
\label{fig3}
\end{figure}

The fidelity with which these operations can be realized as presented here are of course subject to limitations. 
Indeed the excitation must be such that higher than first order corrections are effectively negligible. Furthermore the 
physical realization of a transmission device of this type is also subject to the precision with which the operations 
described above can be realized. This concerns essentially the mechanical and thermal fluctuations of the external field
applied at site $N$.

\subsection {Finite sequences of excitations}
In practical applications the excitation to be transmitted from site $N$ to some site $m$ through 
the system might be repeated a finite number of times starting from an initial time $t=0$. Consider 
the case where a set of $n$ excitations are sent from site $N$ over regular intervals of time $t_{0}\gg \tau_{H}$.
For times $t>nt_{0}$ there are two contributions to $\langle S^{z}_{m}(t)\rangle^{(1)}$,
$\langle S^{z}_{m}(t)\rangle^{(1)} = \langle S^{z}_{m}(t)\rangle^{(1a)} + \langle S^{z}_{m}(t)\rangle^{(1b)}$ 
which read

\begin{figure}
\includegraphics[scale=0.3]{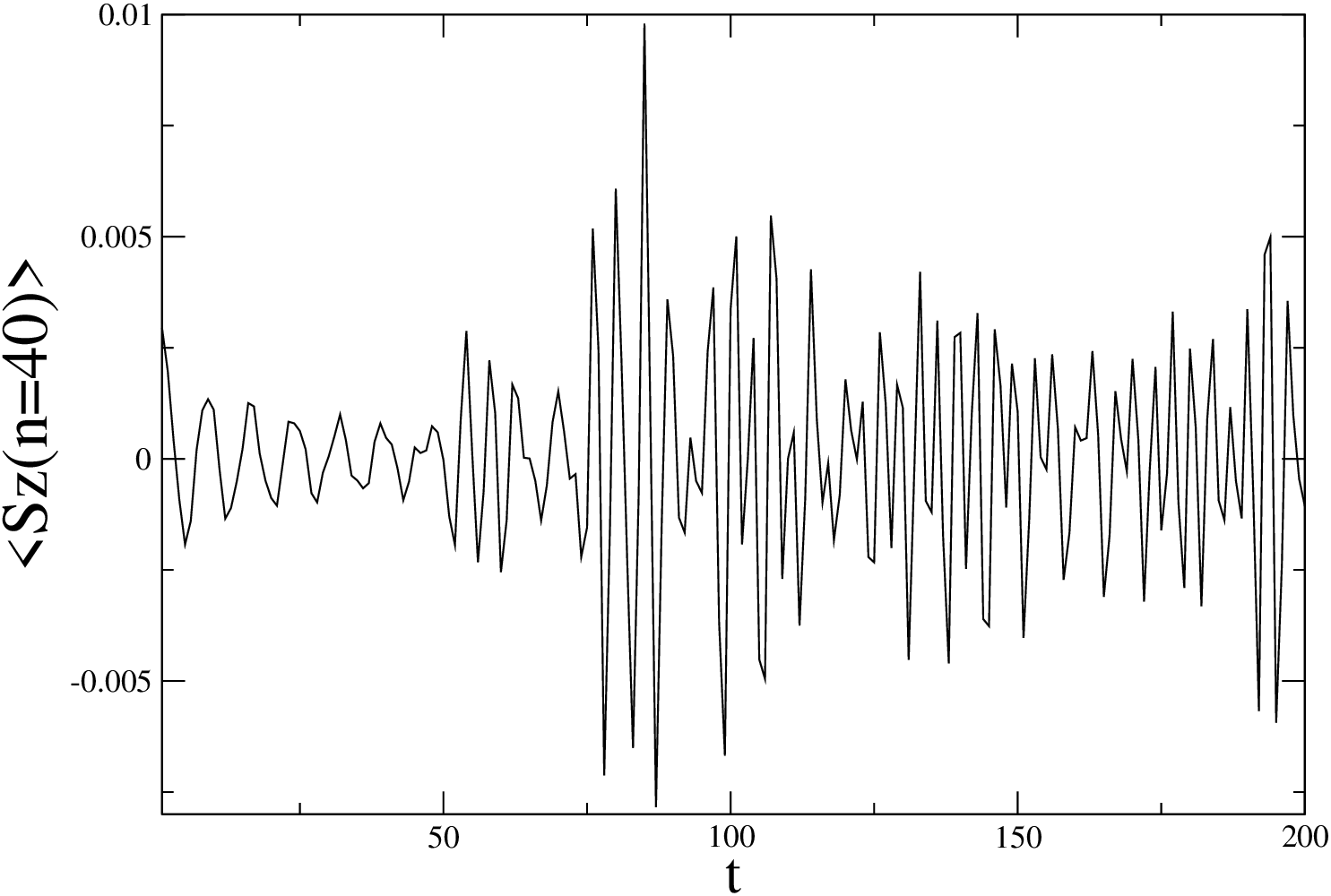}
\caption{Time dependence of the z-component of the spin $\langle S^{z}_{n}(t)\rangle^{(12)}$ at site n=40.}
\label{fig4}
\end{figure}

\begin{eqnarray}
\langle S^{z)}_{m}(t)\rangle^{(1a)}=\sum_{p=0}^{n}\sum_{k,l}\frac{1}{D(k,l)}[\omega_{(kl)} 
\notag \\
(1-\cos(\omega_{(kl)}(t-pt_{0}))\exp(-(t-pt_{0})))-\frac{1}{\tau_{H}}
\notag \\
\sin(\omega_{(kl)}(t-pt_{0}))\exp(-(t-pt_{0}))]
\label{eq14}   
\end{eqnarray}          
where $D(k,l)= [1/\tau_{H}^{2}+\omega_{(kl)}^{2}]$ and 
 
\begin{eqnarray}
\notag \\
\langle S^{z}_{m}(t)\rangle^{(1b)}=\frac{1}{2}\sum_{k,l}v_{2}^{(k,l)}
\notag \\
S_{n}(k,l)[A_{kl}[(\sin(\varphi_{n,m;kl}^{+})+\sin(\varphi_{n,m;kl}^{-})]+
\notag \\ 
+C_{kl}[(\cos(\varphi_{n,m;kl}^{+})+\cos(\varphi_{n,m;kl}^{-})]]
\label{eq15}   
\end{eqnarray}
with

\begin{eqnarray}
S_{n}(k,l) =\frac{\sin[\omega_{(kl)}(n+1)/2t_{0}]}{\sin[\omega_{(kl)}/2t_{0}]}
\label{eq16}   
\end{eqnarray} 
and

\begin{eqnarray}
\varphi_{n,m;kl}^{+}=\omega_{(kl)}(t-nt_{0}/2)+m K(k,l)
\notag \\
\varphi_{n,m;kl}^{-}=\omega_{(kl)}(t-nt_{0}/2)-m K(k,l)
\label{eq17}   
\end{eqnarray} 

The observable $\langle S^{z}_{m}(t)\rangle^{(1)}$ is known at each time $t$ and can in principle be renormalized to 
a fixed value as discussed above in the case of a single excitation.

\section {Summary and outlook}

The propagation of a local perturbation generated on a quantum spin chain at temperature $T=0$ has been investigated. 
The perturbation generated by an external magnetic field starts at an initial time $t=0$ at some local spin site of a $1d$ 
chain and affects the transverse component of the other spins. The conditions imposed on the relaxation time of the external 
field allow a perturbative treatment. 

The action of the external field induces time-dependent oscillations of the transverse components along the spin sites of the 
chain which behaves like a ring due to the imposed boundary conditions. The oscillations of the spin components show the complex behaviour of classical traveling waves with a large number of harmonic modes due to the spectral properties of the physical 
system on which the perturbation propagates. The case where finite sequences of perturbations are locally emitted at fixed time 
intervals of equal length has been worked out.

\begin{figure}
\includegraphics[scale=0.30]{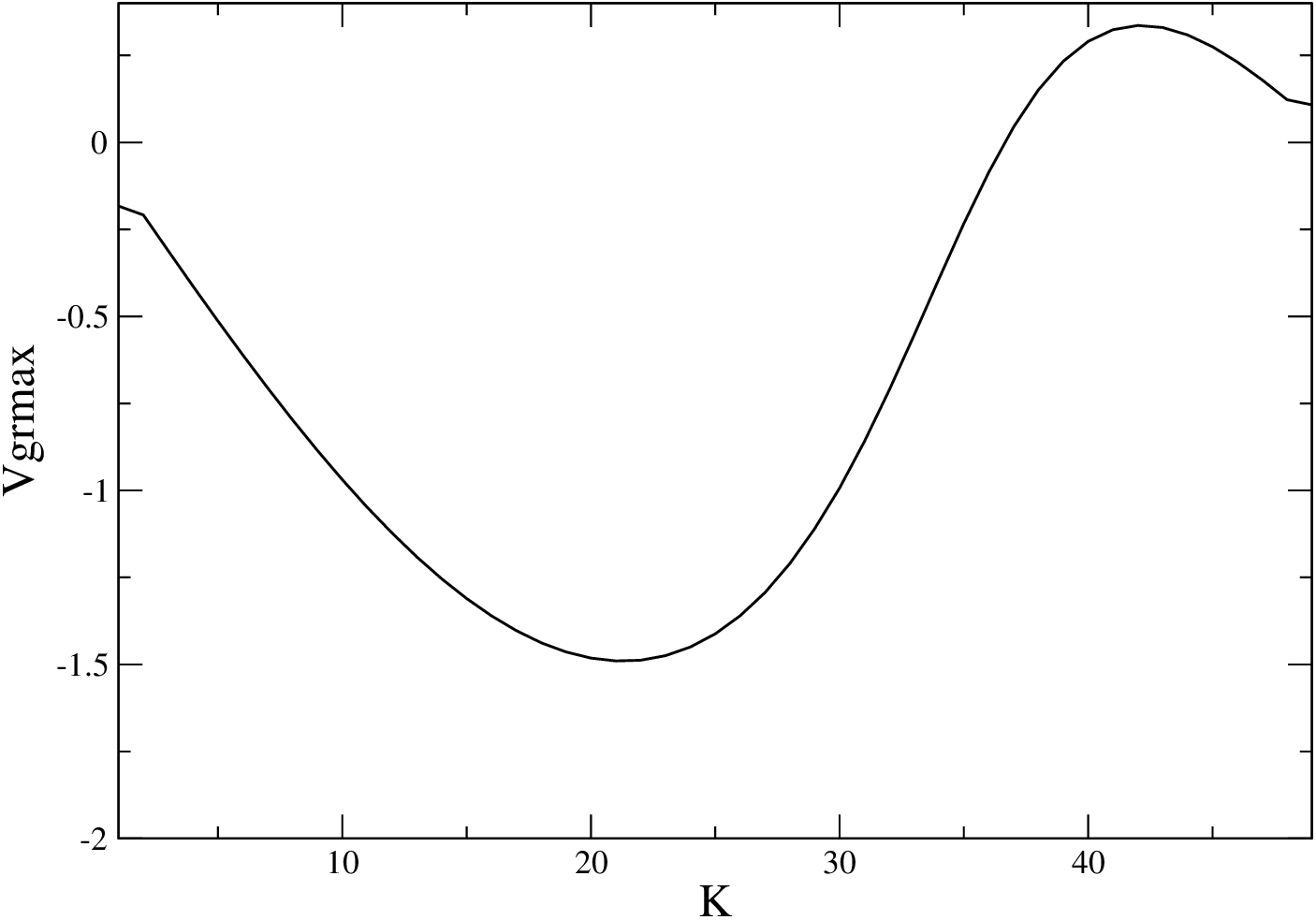}
\caption{Maximum group velocity as a function of K(k,l) for a chain with N=100 and the parameters given in the text.}
\label{fig5}
\end{figure}

In order to characterize the propagation of the perturbation different group velocities have been introduced. These 
velocities show a strong dispersive behaviour far from sharp wave packets. They reflect the complex nature of the strongly 
coupled spins in the chain. It comes out that these velocities are sizably smaller than the Lieb-Robinson propagation velocity 
bound defined in~\cite{lr3,lr4}. Reasons for this have been presented and discussed in section $4$. 

The chain undergoes a phase transition for a specific value $h_{c}$ of the internal magnetic field which acts on the transverse 
components of the spins. The propagating wave is not qualitatively influenced by this property. It is due to the fact that the propagation concerns individual spins and the external field does not affect the global behaviour of the system. This point is 
analysed in section $4$.

The way in which the value of the transverse component of a spin at one site can be retrieved at another site has 
been investigated under ideal noiseless conditions. A more realistic approach should include the presence of noise 
which can be induced by initial thermal effects and fluctuations of the external magnetic field acting on the system. 
It would also be of interest to consider the case of a chain coupled to a heat bath at temperature $T \ne 0$ at time 
$t=0$ which would introduce the action of excited states of the chain.
 
The present study concerns systems which are influenced by a field which acts over a small amount of time and cannot
produce a reversal of the spins. The extension of the present investigation to chains excited by strong fields able to 
affect sensibly or even to reverse the transverse spin direction would request a non-perturbative treatment. Such a case 
needs further developments.

Finally the choice of even boundary conditions on the end sites of the chain closes it in practice. The behaviour of the 
propagation in an open chain may lead to a different behaviour of the perturbation. In the prospect of practical applications 
it would be particularly interesting to investigate this point.

\vskip 0.5cm

Thanks are due to J.-Y. Fortin for a careful reading of the manuscript.\\

{\bf\Large{Appendix A: explicit expressions of the single particle amplitudes}}\\

The energy spectrum of the $1d$ XY chain of length $N$ reduces to a product of single particle energies with $4$ 
states for a given momentum $(k=[1,N/2])$ in Fourier space~\cite{bar}. The local matrix for fixed $k$ can be analytically 
diagonalized leading to the eigenvalues given by the expressions shown in Eq.(3).
The amplitudes of the eigenvectors corresponding to the states $1$ and $2$ are given by
\begin{eqnarray}
\alpha^{k}_{1}&=&(\epsilon^{(1)}_{k}-(2\cos(\phi_{k}-h_0))/N_{(1)}
\notag \\
\alpha^{k}_{2}&=&(-i\delta_{p}/2)/N_{(1)}
\notag\\
\beta^{k}_{1}&=&(i\delta_{p}/2)/N_{(2)}
\notag\\
\beta^{k}_{2}&=&(\epsilon^{(2)}_{k}-h_{0})/N_{(2)}
\label{eq17}   
\end{eqnarray} 
with $\delta_{p}=-2\gamma \sin(\phi_{k})$ and the normalization factors

\begin{figure}
\includegraphics[scale=0.30]{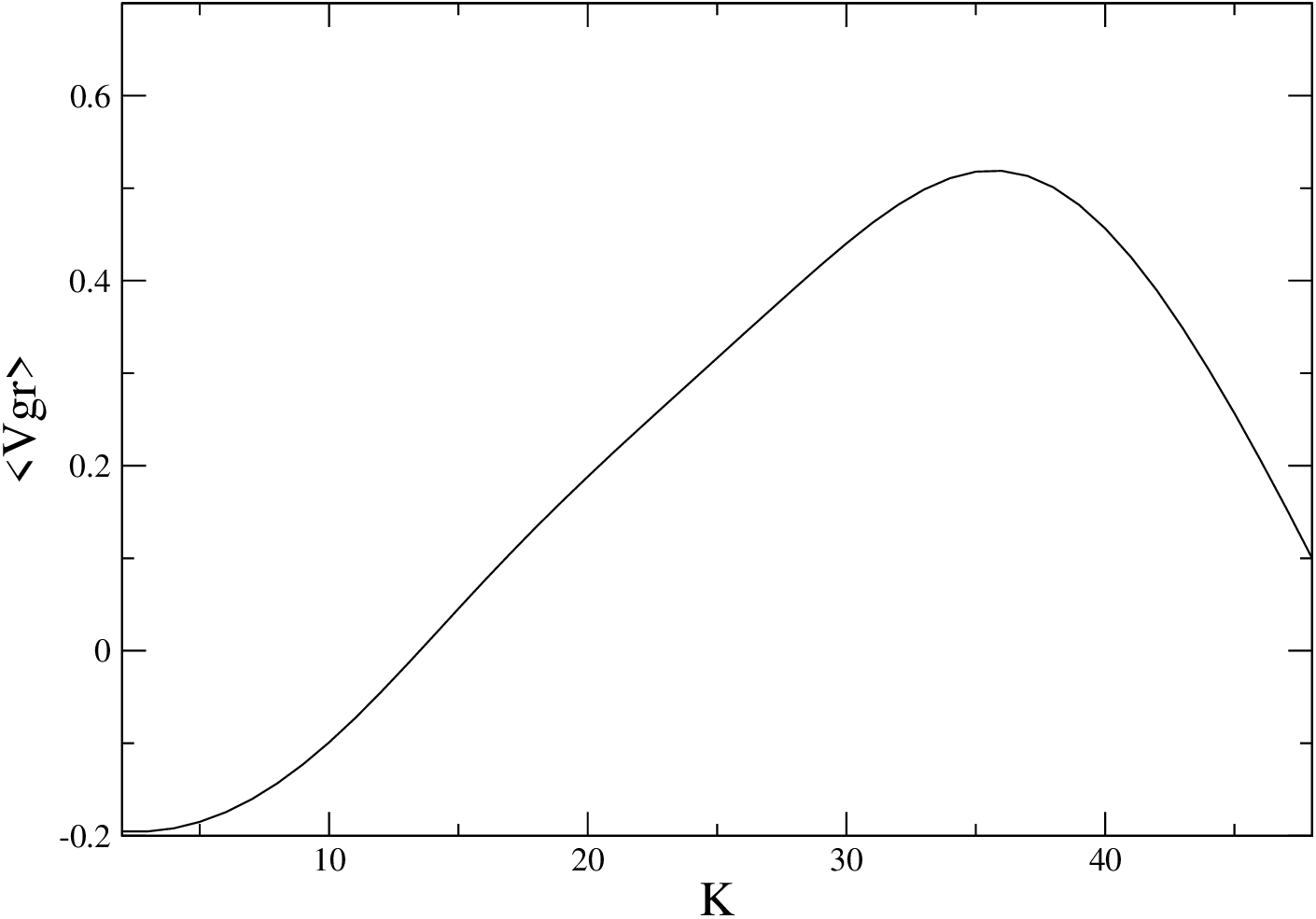}
\caption{Average group velocity as a function of K(k,l) for a chain with N=100 and the parameters given in the text.}
\label{fig6}
\end{figure}

\begin{eqnarray}
N_{(1)}= [1/4 \delta_{p}^2+(\epsilon^{(1)}_{k}-(2\cos\phi_{k}-h_0)^2]^{1/2}
\notag\\
N_{(2)}= [1/4 \delta_{p}^2+(\epsilon^{(2)}_{k}-h_0)^2]^{1/2}
\label{eq18}  
\end{eqnarray} 
\\
\vskip 0.2cm
{\bf\Large{Appendix B: matrix elements and amplitudes entering $\langle S^{z(1)}_{j}(t)\rangle$}}\\

The transition matrix elements $v_{2}^{(k,l)}$ of the first order contributions to $\langle S^{z(1)}_{J}(t)\rangle$
are given by 

\begin{eqnarray}
v_{2}^{(k,k)} = \langle \Psi_{0}(k)|\hat V| \Psi_{i}(k)\rangle
\notag\\
v_{2}^{(k,l)} = \langle \Psi_{0}(k)|\hat V| \Psi_{ij}(kl)\rangle
\label{eq19}  
\end{eqnarray}  
where $|\Psi_{0}(k)\rangle$, $|\Psi_{i}(k)\rangle$, $|\Psi_{ij}(kl)\rangle$ are the single state 
eigenfunctions at sites $(k,l)$ in Fourier space in the ground state $|0>$ and excited states $|i>$ and $|j>$, 
$\hat V$ the time-independent part of the diagonal and non-diagonal elements of the perturbation interaction 
$\hat V(t)= h_{1}\exp(-t/\tau_{H})c^{+}_{N}c_{N}$ expressed in the second quantization formalism in ordinary 
space~\cite{bar}.

The amplitudes $A_{kl}$ and $B_{kl}$ depend on the structure of the chain and the applied magnetic field 
at site $N$. They read:
\begin{eqnarray}
A_{kl}=\frac{1/\tau_{H}}{1/\tau_{H}^2 + (\Delta_{(k,l)}^{(i,j)} \epsilon)^2}
\notag\\
B_{kl}=-\frac{\Delta_{(k,l)}^{(i,j)} \epsilon}{1/\tau_{H}^2 + (\Delta_{(k,l)}^{(i,j)} \epsilon)^2}
\label{eq20}   
\end{eqnarray} 
where $\Delta_{(k,l)}^{(i,j)} \epsilon=\epsilon_{k}^{(i)}+\epsilon_{l}^{(j)}-
\epsilon_{k}^{(1)}-\epsilon_{l}^{(1)}$, $i$ and $j$ corresponding to excited states.\\

{\bf\Large{Appendix C: second order contributions to the tranverse component of spin n}}\\

The order of magnitude of the matrix elements given by the coefficients linked to the time-dependent 
part of the different contributions of second order are given by the strength factors $S_{tr}$
\begin{eqnarray}
S_{1} = h_{1}^{2} \tau_{H}^{2}
\notag\\
S_{2} = h_{1}^2  (1/D^2)
\notag\\
S_{3} = h_{1}^2/(\tau_{H} D)
\notag\\
D \sim (1/\tau_{H}^2 + \Delta \epsilon^2)
\label{eq21}   
\end{eqnarray} 
where $\Delta \epsilon \sim \epsilon^{(i)}-\epsilon^{(j)}$, $(i,j)$ corresponding to single particle
ground or excited states. The single particle state energies are of the order of unity. In the numerical 
applications $h_{1}$ is chosen to be of the same order of magnitude and $\tau_{H}$ is four order of magnitude 
smaller. As a consequence the second order contributions to $\langle S^{z}_{j}\rangle$ are two to three orders 
of magnitude smaller than the first order ones. This justifies their neglect in the numerical estimates.
 
\vskip 0.3cm

{\bf\Large{Appendix D: estimate of the Lieb-Robinson bound for the chain.}}

\vskip 0.3cm
The expression of the velocity reads

\begin{eqnarray}
v_{a}=\inf_{a} C_{a}\|\Phi_{a}\|/a
\end{eqnarray}
Here $a$ is a real positive number, $C_{a}$ depends on the distances between discrete sites $(x,y,z)$ of the system and 
\begin{eqnarray}
\|\Phi_{a}\|=\|H\|/D_{a}(x,y)
\end{eqnarray}
where $\|H\|$ is the norm of the Hamiltonian matrix and $D_{a}(x,y)$ a measure of the distance between sites $x$ and $y$ multiplied 
by an exponential factor $exp(-a|x-y|)$~\cite{lr2,lr3}.
 
Working out this expression leads to the expression

\begin{eqnarray}
v_{a}=\inf_{a}  N/2 \exp(aN)\Phi(a=1)/aN
\end{eqnarray}
which shows an extremum for $a=1/N$. This extremum is a minimum. The final result is given by $v_{LR}$ in section $3$.

\end{document}